\documentclass[runningheads]{llncs}

\usepackage{amsmath}
\usepackage{amssymb}
\usepackage{gensymb}

\usepackage{svg}
\usepackage{hyperref}
\usepackage{cleveref}

\usepackage[T1]{fontenc}

\usepackage{graphicx}
\usepackage{xcolor}

\newcommand{\R}{\mathbb{R}}
\newcommand{\norm}[1]{\left\Vert #1 \right\Vert}

\newcommand{\Tree}{\mathcal{T}}
\newcommand{\pos}[1]{#1}
\newcommand{\emb}[1]{\overline{#1}}
\newcommand{\ang}[1]{\theta(#1)}

\DeclareUnicodeCharacter{03C6}{$\varphi$}

\begin{document}
\title{Untangling Vascular Trees for Surgery and Interventional Radiology}

\author{Guillaume Houry \inst{1}\orcidID{0000-0002-1918-0545} \and
Tom Boeken\inst{1, 2}\orcidID{0000-0003-2990-5180} \and
Stéphanie Allassonnière\inst{1}\orcidID{0000-0002-5692-4945}
\and \\
Jean Feydy\inst{1}\orcidID{0000-0001-6049-563X} }

\authorrunning{G. Houry et al.}

\institute{Inria, Université Paris Cité, Inserm, HeKA, F-75015 Paris, France \\
\and
Department of Vascular and Oncological Interventional Radiology, \\
Hôpital Européen Georges Pompidou, AP-HP, 75015 Paris, France \\
\email{jean.feydy@inria.fr}}
\maketitle           

\begin{abstract}
The diffusion of minimally invasive, endovascular interventions motivates the development of visualization methods for complex vascular networks. 
We propose a planar representation of blood vessel trees which preserves the properties that are most relevant to catheter navigation: topology, length and curvature. 
Taking as input a three-dimensional digital angiography, our algorithm produces a faithful two-dimensional map of the patient’s vessels within a few seconds. 
To this end, we propose optimized implementations of standard morphological filters and a new recursive embedding algorithm that preserves the global orientation of the vascular network.
We showcase our method on peroperative images of the brain, pelvic and knee artery networks. On the clinical side, our method simplifies the choice of devices prior to and during the intervention. This lowers the risk of failure during navigation or device deployment and may help to reduce the gap between expert and common intervention centers. From a research perspective, our method simulates the cadaveric display of artery trees from anatomical dissections. This opens the door to large population studies on the branching patterns and tortuosity of fine human blood vessels. Our code is released under the permissive MIT license as part of the \href{https://scikit-shapes.github.io}{scikit-shapes}  Python library\footnote{A demonstration of our code is available in the following folder: \\ \href{https://github.com/scikit-shapes/scikit-shapes/tree/main/arteries-atlas}{\texttt{github.com/scikit-shapes/scikit-shapes/tree/main/arteries-atlas}}.}.

\keywords{Vascular trees  \and Visualization \and Computational anatomy \and Computer-assisted intervention \and Interventional radiology.}
\end{abstract}
\begin{figure}[b!]
\centering
\includegraphics[width=0.99\textwidth]{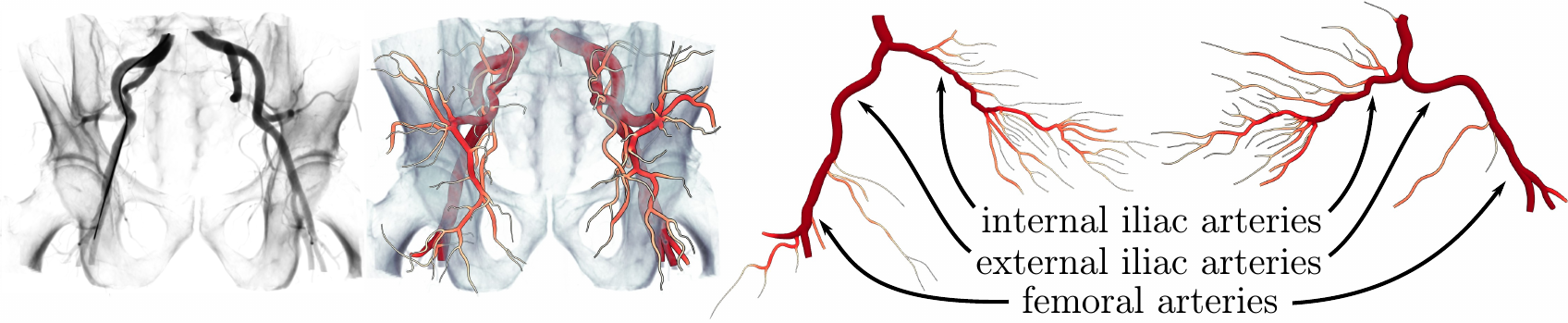} 
\caption{\textit{Left:}~Our method takes as input a three-dimensional angiogram, such as this scan of the pelvic region.
\textit{Middle:}~First, we apply fast morphological filters to segment the blood vessels.
\textit{Right:}~Then, our novel tree embedding algorithm creates a planar vessel map that is optimized for the planning of endovascular interventions.} \label{fig:pelvis}
\end{figure}

\section{Introduction and Related Works}
\label{section:intro}

\subsubsection{Endovascular Interventions.}
Unfolding curved anatomical structures for visualization on two-dimensional displays is a fundamental task in medical imaging, with applications ranging from the detection of rib fractures \cite{ringl2015ribs} to the planning of tumor ablations \cite{lichtenberg2019distance}. Meanwhile, the emergence of endovascular interventions has sparked increasing interest in the detailed mapping of complex arterial networks, understood as modern highways for surgeons, cardiologists and interventional radiologists \cite{acebes2024centerline,naeem2024trexplorer,sinha2024trind,wittmann2024simulation}.
Our work lies at the intersection of these developments: we focus on generating planar representations of intricate vascular structures to facilitate catheter navigation during interventions. Specifically, we aim to embed vascular networks in 2D while preserving key geometric features for effective endovascular guidance: the vessel lengths and curvatures.

\subsubsection{Vessel Tree Analysis and Visualization.}
The creation of blood vessel maps is an active field of research; we refer to \cite{eulzer2022vessel} for a recent overview.
Our work places a strong emphasis on vessel lengths and curvatures, features that are critical for catheter navigation but often overlooked in existing methods.
This focus presents unique challenges: while efficient methods can embed spatial graphs in the plane without intersections \cite{lichtenberg2019distance,pandey2019cerebrovis}, doing so while preserving the length and tortuosity of curves is not straightforward.

Several techniques have been proposed to produce geometry-accurate planar representations of vascular networks.
Curved planar reformations \cite{kanitsar2002cpr,kanitsar2003advanced,won2009uncluttered} generate 2D slices from volumetric images via non-linear projection along vessel centerlines, preserving local geometry and surrounding context. However, they require complex decluttering and cannot guarantee intersection-free layouts.
Alternatively, conformal mappings \cite{zhu2002conformal} flatten the blood vessels while preserving boundary properties. 
Yet, these approaches neglect the global vascular structure, making it unsuited to our purpose.
A last approach, arguably the closest to our objective, consists in building the embedding by joining and untangling multiple local linear projections of the vessel tree \cite{hu2022geometry,marino2015planar}, offering a good balance between geometric fidelity and global readability.

Unfortunately, these techniques are not fully adapted to our problem.
They usually neglect the branching angles at the vessel junctions, which are of critical importance in endovascular navigation.
More importantly, current methods do not scale well with large vascular systems: for instance, \cite{won2009uncluttered} requires 12 minutes to process a tree with 72 branches, while \cite{hu2022geometry,marino2015planar} needs several minutes for networks with a few thousand nodes.
In contrast, the vascular trees we analyze are significantly more complex, featuring dense branching patterns and intricate topologies comprising over 10,000 nodes and more than 500 branches.

\subsubsection{Contributions.} 
To tackle this challenge, we propose an algorithm that produces clear visualizations of complex vascular trees while preserving vessel curvature as accurately as possible.
Our method generates qualitatively satisfying layouts in a few seconds, even in our most challenging cases.
The resulting embeddings contain no intersections, exactly preserve vessel lengths, and maintain the correct angles at the majority of branching points.
Finally, we provide a portable code that implements our full pipeline as part of the scikit-shapes Python library (\href{https://github.com/scikit-shapes/scikit-shapes/tree/main/arteries-atlas}{\texttt{github.com/scikit-shapes/scikit-shapes/tree/main/arteries-atlas}}).

\section{Methods}

Our method takes as input an acyclic vascular network understood as a tree $\Tree = (V, E)$ whose root node $v_{\text{root}}$ corresponds to the blood source. 
We identify each vertex $ {v\in V}$ with its vector of 3D coordinates $v\in \mathbb{R}^3$.
We also require an estimate of the local vessel radius at every vertex.

Because most vertices have exactly two neighbors, we can decompose the tree into a set of contiguous branches $B = (b_1, \dots, b_N)$ that are topologically equivalent to segments.
This induces a coarse-grained directed tree structure $\Tree_{\text{coarse}} = (J, B)$ where $J$ is the subset of vertices with a degree different from two -- namely, the root, junctions, and leaves. These vertices are connected by the branches in $B$.
Our method consists of two main steps:
\begin{enumerate}
    \item A recursive planar embedding of the branches of $\Tree_{\text{coarse}}$ without intersections that prioritizes the preservation of the angles at junctions;
    \item A refinement step using a force-directed scheme to enhance the final layout.
\end{enumerate}
This section provides a detailed overview of our method. We begin by describing the data and preprocessing steps, followed by a formal definition of the planar embedding problem using an angular reparametrization of vertex positions.
We then use this angular representation to define a target curvature at each tree node and conclude with a description of the two steps of our embedding algorithm.

\subsubsection{Data and Preprocessing.}
We work with three-dimensional angiograms acquired via 3D X-ray imaging -- more precisely, a Philips Azurion system following a Cone Beam CT (CBCT) protocol.
Voxel values are expressed in Hounsfield units and blood vessels are highlighted using a contrast agent.
Motivated by applications in stroke surgery, our primary focus is on cerebral vascular networks.
The injection protocol selectively highlights one hemisphere of the brain arterial network, allowing us to assume a tree-like topology. To showcase the generalizability of our method, we also apply it to arterial networks around the pelvis and knee (see \cref{fig:pelvis,fig:knee}). This study was approved by two institutional review boards (Registre général des traitements de l’Assistance Publique–Hôpitaux no. 20220128085623 and CERAR Institutional Review Board 00010254-2022-025) in accordance with the Declaration of Helsinki.

\begin{figure}[p!]
\centering
\includegraphics[width=\textwidth]{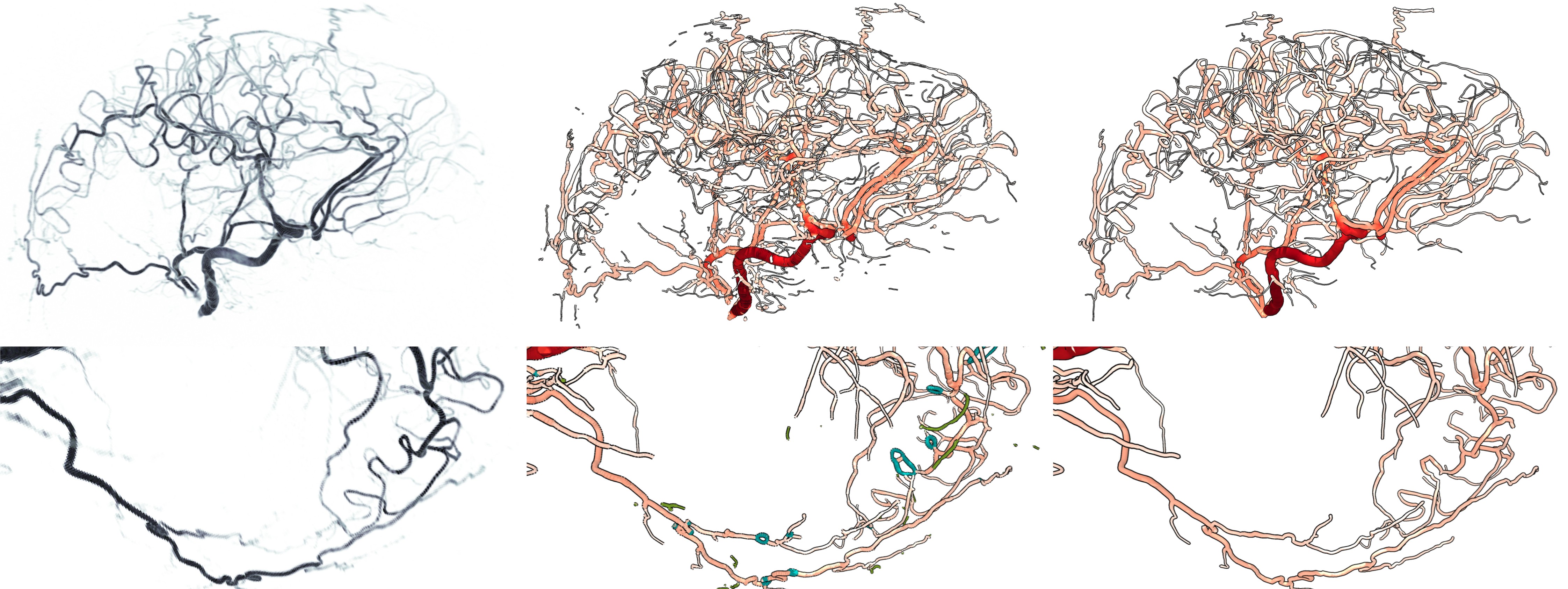} 
\caption{\textit{Left:} Three-dimensional cerebral angiogram. \textit{Middle:} The segmented graph. Shades of red encode the estimated vessels radius, blue edges correspond to cycles and green edges belong to small disconnected components. \textit{Right:} The final vessel tree.} \label{fig:initial_graph}
\end{figure}

As an initial processing step, we convert the 3D images into graphs using efficient implementations of classical image processing techniques: hysteresis thresholding \cite{canny1986computational}, signed distance transform, Frangi filtering \cite{frangi1998multiscale}, and skeletonization \cite{palagyi19983d}. To correct topological artifacts such as small cycles or multiple disconnected components, we compute the minimum spanning tree of the graph \cite{kruskal1956shortest,2020SciPy-NMeth}.
We illustrate this learning-free pipeline in \cref{fig:initial_graph}.
Although not state-of-the-art, this simple method produces satisfying segmentations for our high-resolution data. Leveraging PyTorch \cite{pytorch} and the Taichi library \cite{taichi}, our code is able to process images of size ${500\times500\times500}$ within seconds.

\subsubsection{Angular Parametrization.}
The planar embedding task consists in associating a 2D position $\emb{v}\in\mathbb{R}^2$ to each vertex ${v\in\mathbb{R}^3}$ of the three-dimensional vascular tree $\Tree$.
In order to guarantees the preservation of vessel lengths, we propose to 
use an intermediate representation
via real-valued angular coordinates $\ang{v}$.
Placing the root node at the origin so that $\emb{v}_\text{root} = (0, 0)$, we compute the node positions $\emb{v}\in\mathbb{R}^2$ from the values of $\ang{v}$ recursively with:
\begin{equation}
    \label{ref:eq_angularparam}
    \text{for all edge } (v \rightarrow v') \in E, \quad \emb{v}' ~=~ \emb{v} + \norm{\pos{v} - \pos{v'}}_{\R^3} \cdot \big( \, \cos \ang{v'}, \, \sin \ang{v'} \, \big)~.
\end{equation}

\subsubsection{Computing Signed Curvatures on the Vessel Tree.}
\begin{figure}[p!]
\centering
\includegraphics[width=\textwidth]{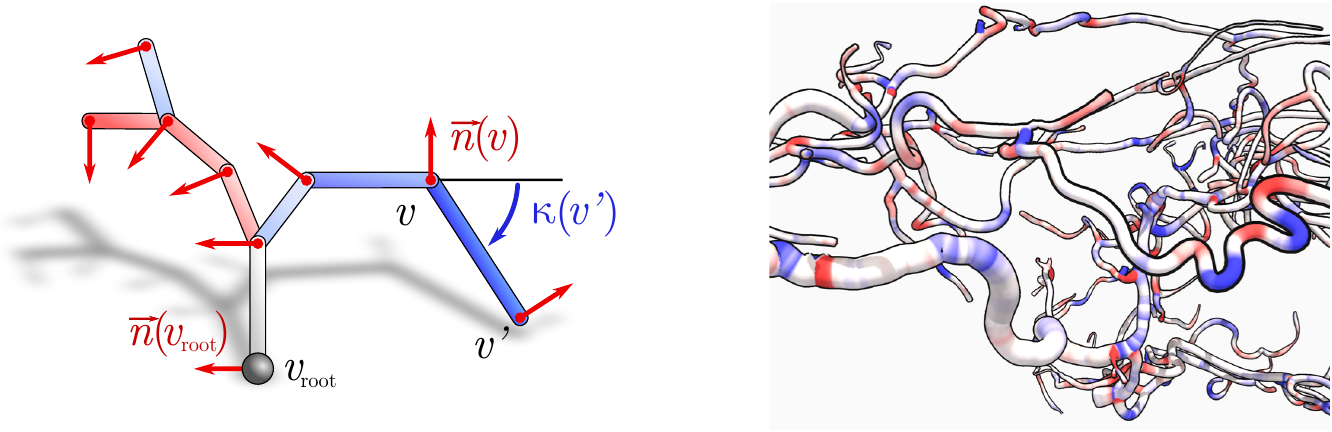} 
\caption{\textit{Left:} 
Flowing from the root node $v_\text{root}$, we define the point curvature
$\kappa(v)$ as the angle between two successive segments of the curve.
We transport a reference normal vector $\overrightarrow{n}(v)$
from $v_\text{root}$ as in \cite{hanson1995parallel}
and use the sign of the dot product
$\overrightarrow{n}(v) \cdot \overrightarrow{vv}{}'$ to orient
the curvature $\kappa(v')$.
\textit{Right:} We display the values of $\kappa(v)$ on a typical example:
red segments are associated to positive values, i.e. left turns in our planar layout; blue segments correspond to negative values, i.e. right turns.} \label{fig:curvature_def}
\end{figure}

Since endovascular navigation is strongly influenced by the angles at vessel junctions and the accumulation of torque along the catheter, it is important to reflect the original 3D vessel curvatures in the 2D embedding.
To this end, we impose a constraint on bending angles: for any three consecutive nodes ${v \rightarrow v' \rightarrow v''}$, the absolute difference ${\vert \theta(v) - \theta(v') \vert}$ should match the angle between the 3D segments $\overrightarrow{vv'}$ and $\overrightarrow{v' v''}$. 
To capture the correct global shape of the vessel, we must also assign the appropriate sign to this difference, indicating a left or right bend in the 2D plane.
Following \cite{hanson1995parallel}, we use parallel transport to propagate a reference normal vector $\overrightarrow{n}(v)$ along the tree $\Tree$ and define the bending direction using the sign of the dot product $\overrightarrow{n}(v)\cdot \overrightarrow{vv'}$ (see \cref{fig:curvature_def}).
This allows us to define a signed angular curvature $\kappa(v)$ at each vertex. Preserving the curvature of blood vessels in the planar visualization is then equivalent to making sure that:
\begin{equation}
    \text{for all edge } (v \rightarrow v') \in E, \quad \ang{v'} ~\simeq~  \ang{v} + \kappa(v')~.
    \label{ref:eq_curvature}
\end{equation}

\subsubsection{Intersection-free Planar Layout.} 
\begin{figure}[p!]
\centering
\includegraphics[width=.8\textwidth]{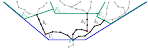} 
\caption{At every new junction $v$, we draw downstream branches $b_1,\dots, b_k$ by integrating eq.~(\ref{ref:eq_curvature}) until colliding with the boundary of $U_v$ or with any other branch (red nodes). We embed the rest of the branch as a line, subdivide the domain $U_v$ in angular sectors $U_1,\dots, U_k$ and proceed iteratively.} \label{fig:recursive_algo}
\end{figure}

In general, the constraint in \cref{ref:eq_curvature} cannot be satisfied simultaneously for all vertices without causing edge intersections. 
As a first step that we illustrate in \cref{fig:recursive_algo},
we compute an embedding that guarantees no intersections and attempts to preserve angle variations around branching points.
To this end, we work with the coarse-grained structure
$\Tree_\text{coarse}$ and recursively tesselate the embedding plane in
polygonal regions $U_v$ that contain the embedding locations
of all vertices downstream of point $v$.

\begin{figure}[t!]
\centering
\includegraphics[width=.75\textwidth]{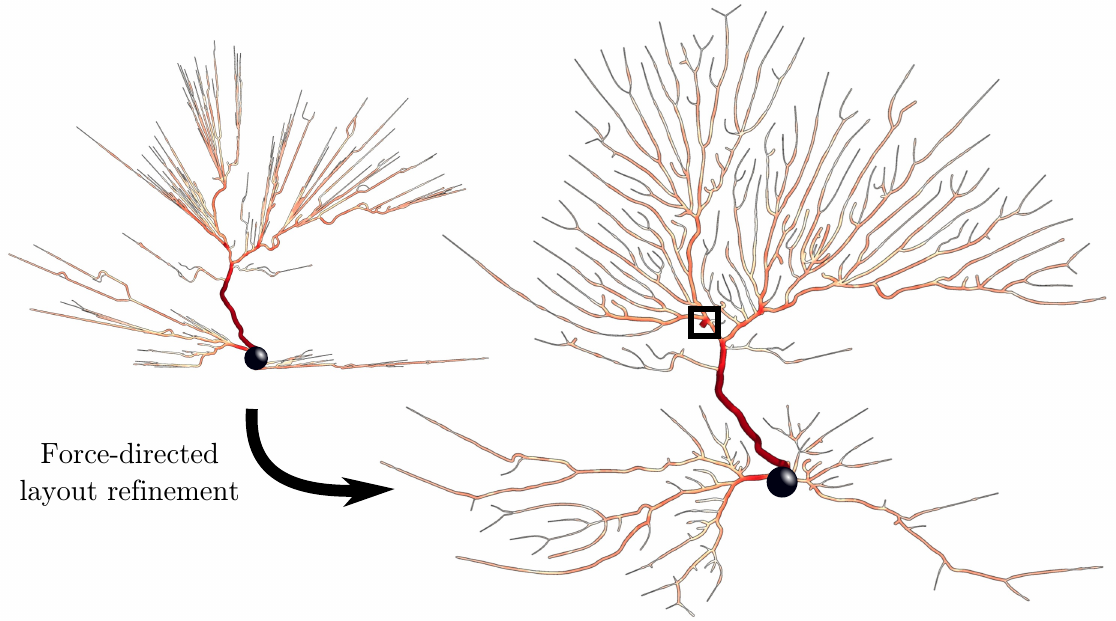} 
\caption{\textit{Left:} Result of the recursive embedding for the brain artery tree of \cref{fig:initial_graph} (running time: 1\,s). \textit{Right:} Final embedding, after 300 steps of iterative refinement (running time: 2\,s). The layout closely mimics pictures of genuine anatomical dissections \cite{kahilogullari2012branching}, with the aneurysm clearly visible (black square). For the sake of clarity, we display both internal (top) and superficial (bottom) arteries, artificially tying them together around the root node (black disk).} \label{fig:illustr_algo}
\end{figure}

First, we associate the upper half-plane $U_{v_\text{root}} = \mathbb{R} \times \mathbb{R}_{\geqslant 0}$
to the root node embedded at the origin point $\emb{v}_\text{root} = (0, 0)$.
Then, working recursively at every junction node $v$,
we embed downstream branches $b_1, \dots, b_k$ by integrating \cref{ref:eq_curvature}. This exactly preserves vessel curvatures until a collision is detected, as highlighted in red in \cref{fig:recursive_algo}. Such collisions can occur either with previously embedded parts of branches $b_1, \dots, b_k$ or with the boundary of the region $U_v$.
When a collision is encountered, we embed the remaining portion of the affected branch as a straight line. The domain $U_v$ is then subdivided into $k$ non-overlapping sectors $U_{v'_1}, \dots, U_{v'_k}$ associated with the endpoints of the branches $b_1, \dots, b_k$. These sectors define feasible regions for the subsequent recursive calls at the child nodes $v'_1, \dots, v'_k$ in $\Tree_\text{coarse}$.
We show a typical output of this recursive embedding method on the left panel of \cref{fig:illustr_algo}.

\subsubsection{Force-directed Layout Refinement.} 
This initial embedding straightens distal vessels. 
To enhance clarity, we denote by $\theta_0(v)$ the angles produced by our recursive embedding algorithm and iteratively apply the following update:
\begin{equation*}
    \theta_{t+1}(v) \gets \theta_t(v) + \lambda_{\text{rep}} \cdot F_{\text{rep}}(v) + \lambda_{\text{bend}} \cdot F_{\text{bend}}(v) + \lambda_{\text{struct}} \cdot F_{\text{struct}}(v)~,
\end{equation*}
where $F_{\text{rep}}$ is a repulsive force between nearby vertices that prevents overlaps, $F_{\text{bend}}$ bends the angles towards their true curvature and $F_{\text{struct}}$ maintains the global structure of the initial embedding.
We use PyTorch to implement the following functions at every node $v$ with parent $p$ and vessel radius $r$:
\begin{equation*}
\begin{split}
    F_{\text{rep}}(v) &= - \frac{\partial}{\partial \theta_t(v)} \sum_{v' \neq v} \left[ \left(  \frac{1}{\norm{\emb{v} - \emb{v}'}^2_{\mathbb{R}^2}}  + \frac{\alpha}{\norm{\emb{v} - \emb{v}'}_{\mathbb{R}^2}} \right) \cdot \exp \left(-\tfrac{1}{\sigma}\norm{\emb{v} - \emb{v}'}_{\mathbb{R}^2}\right) \right]~, \\
    F_{\text{bend}}(v) &= r^{\mu} \cdot \left[ \kappa(v) - (\theta_t(v) - \theta_t(p)) \right]~, \quad F_{\text{struct}}(v) = \theta_0(v) - \theta_t(v)~.
\end{split}
\end{equation*}
$F_{\text{rep}}$ is the sum of two forces operating at different scales: the first prevents vessel crossings while the second spreads the branches to enhance their readability. The exponential factor limits the repulsion range for numerical stability.  The term  $r^\mu$ in $F_{\text{bend}}$ prioritizes the curvature of the largest vessels over the finest ones.  A typical result is shown on the right panel of \cref{fig:illustr_algo} with $\lambda_{\text{rep}} = 25$, $\lambda_{\text{bend}} = 0.4$, $\lambda_{\text{struct}} = 5\cdot10^{-5}$, $\alpha = 0.1$, $\sigma = 200$ and $\mu = 1$. 
These parameters influence the embedding style and should be tuned according to the anatomical region’s characteristics, such as average vessel length and density of ramifications.

\section{Results and Conclusion}
\label{section:results}

\subsubsection{Surgical Planning and Anatomical Comparisons.}
\begin{figure}[p!]
\centering
\includegraphics[width=.9\textwidth]{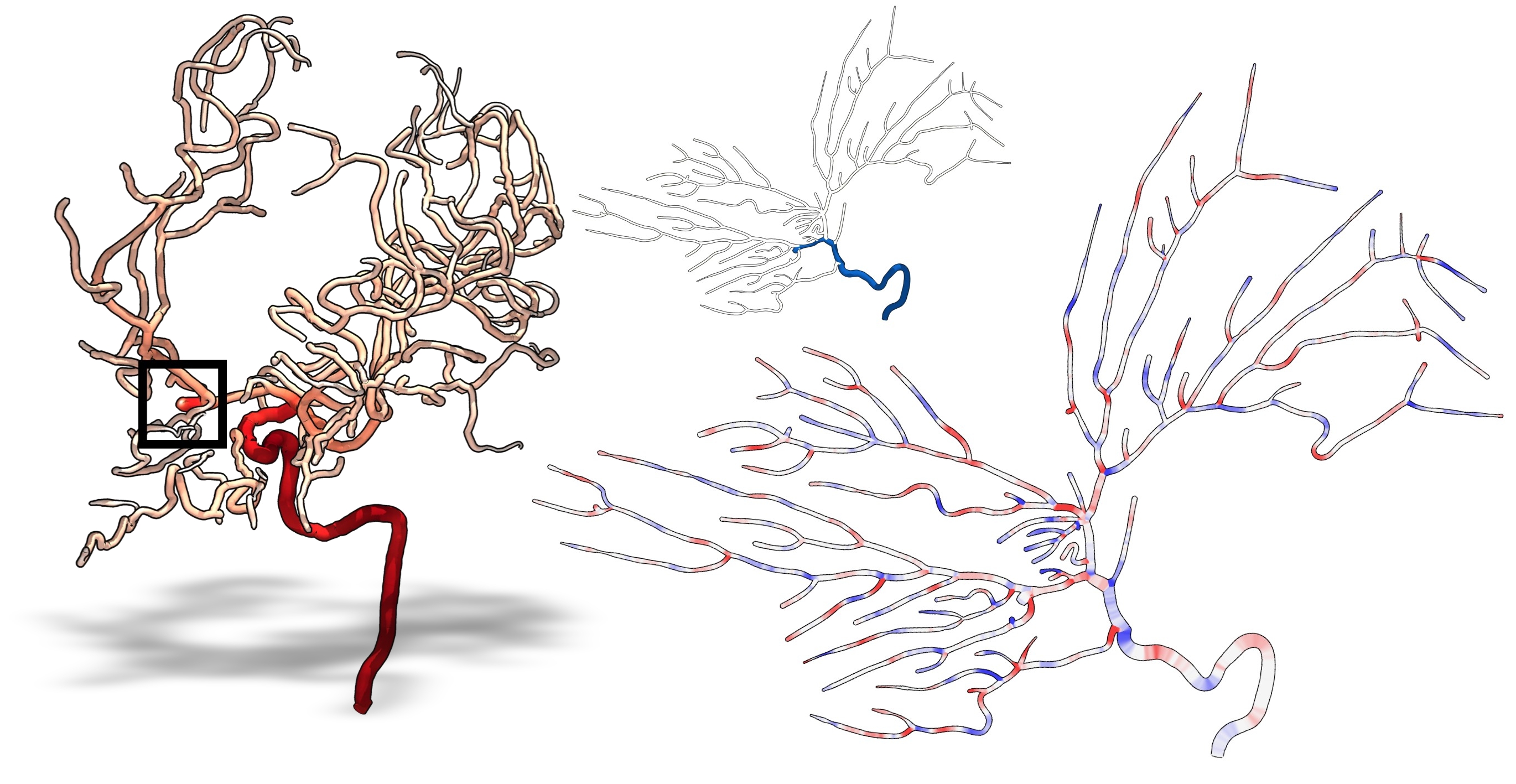} 
\caption{\textit{Left:} Segmented brain artery tree in 3D, with an aneurysm (black square). \textit{Top:} Path from the root node to the aneurysm. \textit{Right:} Planar layout, colored by curvature.} \label{fig:result_intervention}
\end{figure}

As illustrated in \cref{fig:result_intervention}, our method turns peroperative three-dimensional angiograms of the brain arterial network into legible planar representations. 
Unlike preexisting approaches, our layout preserves junction angles, vessel lengths and curvatures. This is especially relevant for interventional radiology; at a glance, a physician can now understand the shape of the path to be followed. This vessel map is thus ideally suited to guide the choice and thermo-forming of a catheter with a suitable tip and stiffness.
Crucially, our implementation is released under the permissive MIT license and runs in less than ten seconds on a standard laptop CPU: this opens the door to its integration in the interventional workflow.

\begin{figure}[p!]
\includegraphics[width=\textwidth]{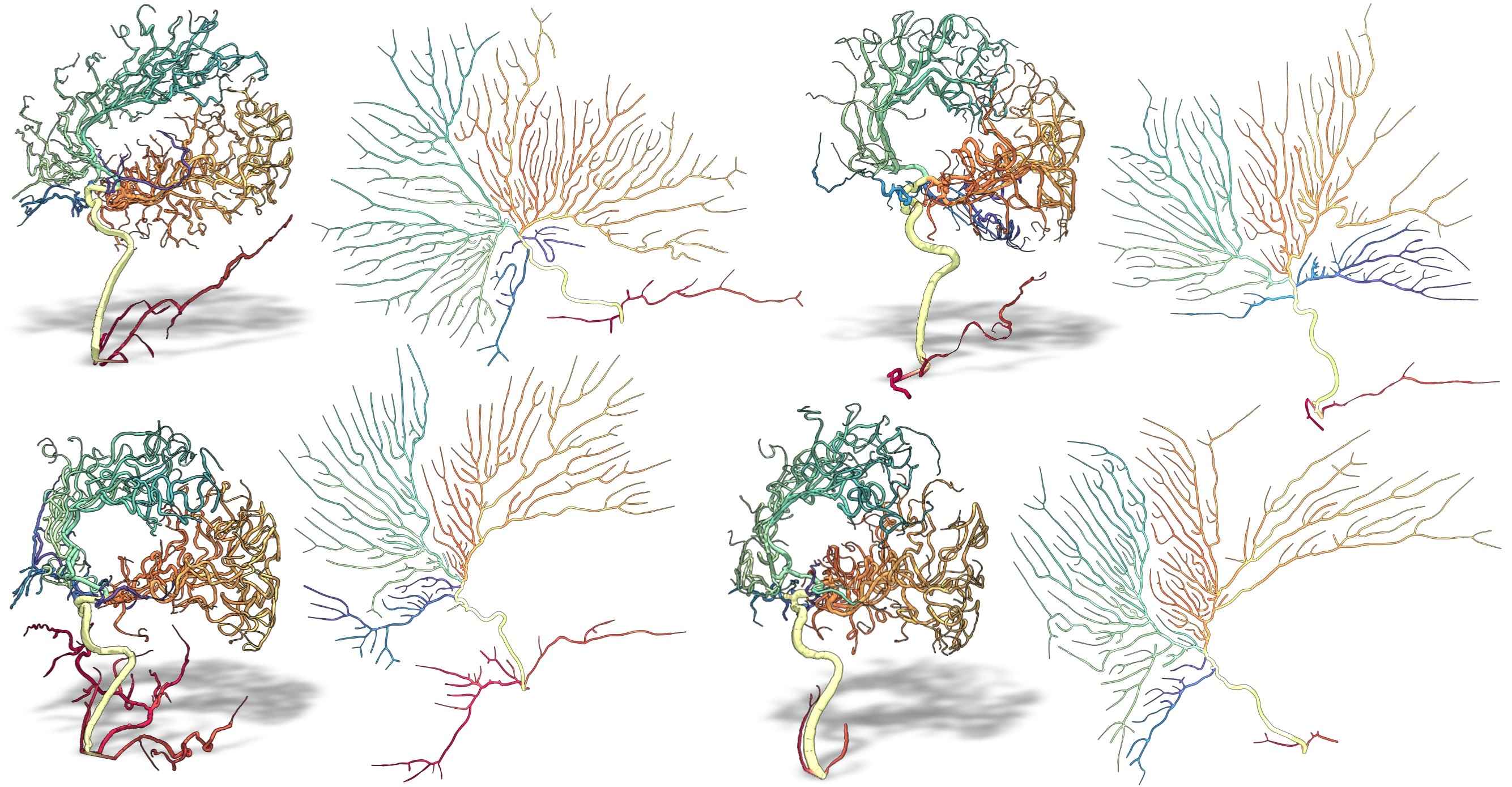} 
\caption{Untangling four cerebral vascular trees that include
the left carotid artery (yellow), anterior cerebral artery (green) and middle cerebral artery (orange).} \label{fig:gallerie_exemples}
\end{figure}

Our method also enables the visualization of detailed images of the cerebral vascular system that may contain very distal vessels.
Due to their intricate geometries, the original three-dimensional volumes are difficult to read even for experts: our planar layouts solve this issue at a minimal computational cost.
Our method is thus similar to Cerebrovis \cite{pandey2019cerebrovis}, but with a focus on geometry instead of pure topology. 
In \cref{fig:gallerie_exemples}, we display the brain artery networks of four patients that underwent surgery for aneurysms: we believe that our vessel map will help physicians formulate novel hypotheses on, for example, risk factors for aneurysms or thromboses, which may then be validated by a morphometric study \cite{guo2022statistical}.


\begin{figure}[p!]
\centering
\includegraphics[width=\textwidth]{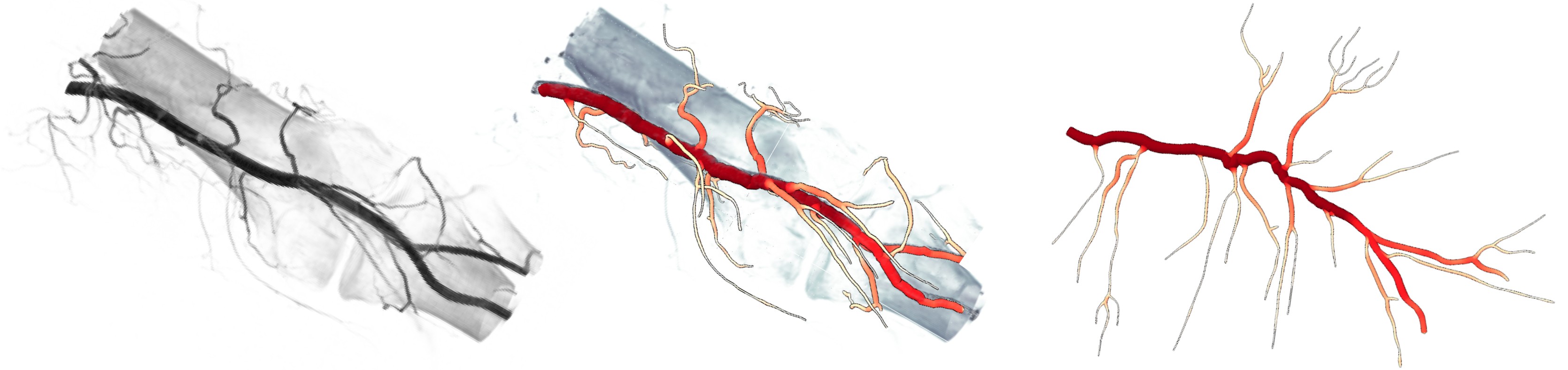} 
\caption{\textit{Left:} Three-dimensional angiography of the popliteal artery across the knee.
\textit{Middle:} Segmented artery tree, colored by the vessel radius.
\textit{Right:} Planar layout.
} \label{fig:knee}
\end{figure}

\subsubsection{Robustness and Quantitative Evaluation.}
Although vessel segmentation is an active research topic \cite{moccia2018blood}, we found that our learning-free pipeline works surprisingly well on peroperative three-dimensional angiograms. 
Notably, applying a vesselness filter to the signed distance function instead of raw intensity values is enough to prevent the ``fusion'' of vessels that run parallel to each other such as the left and right anterior cerebral arteries. This is testament to the superior quality of data that is acquired in an interventional context, with high resolution devices and no motion artifacts.
Although our work was first motivated by cerebral interventions, we stress that our method performs just as well on other anatomical regions such as the pelvis of \cref{fig:pelvis} and the knee of \cref{fig:knee}.

In the examples of \cref{fig:illustr_algo,fig:result_intervention,fig:gallerie_exemples,fig:knee}, our method preserves, within $\pm 10 \degree$, $75 \%$ to $90\%$ of junction angles between vessels with radius larger than 2 mm. This proportion drops to between $45\%$ and $55 \%$ for smaller vessels, which are necessarily compressed in the 2D embedding.
Likewise, the curvature constraint of \cref{ref:eq_curvature}
is satisfied up to $2.5 \degree$ on average for large vessels
and $4.0 \degree$ for smaller vessels.
This should be compared with the standard deviation of $5.2 \degree$ for the bending angles $\kappa(v')$.
These results highlight our method’s ability to preserve the geometry of major vessels while tolerating some loss of accuracy in finer vessels whose detailed shapes are less clinically relevant.

\subsubsection{Limitations and Future Works.}
Going forward, we believe that our method could benefit from three major improvements.
First, going beyond scalar radii to handle non-circular vessel sections would be desirable, especially in the context of interventions on aneurysms and thromboses.
Second, we intend to combine our embedding algorithm with state-of-the-art segmentation networks to handle more common but challenging data, such as routinely acquired magnetic resonance images.
Finally, extending our method to arterial networks that contain cycles (such as the circle of Willis) would be a significant breakthrough; this is a fundamental research problem, for which we may leverage recent insights from the computer graphics literature \cite{Minarcik2024Minkowski}.

\begin{credits}
\subsubsection{\ackname} This study was funded by the France 2030 program managed by the Agence Nationale de la Recherche (ANR-23-IACL-0008, PR[AI]RIE-PSAI).

\subsubsection{\discintname}
The authors have no competing interests to declare that are
relevant to the content of this article.
\end{credits}

\bibliographystyle{splncs04}
\bibliography{bibliography}

\end{document}